\documentclass[a4paper,11pt]{article}
\pdfoutput=1 

\usepackage{jheppub} 

\usepackage[T1]{fontenc} 

\usepackage{slashed} 

\def\sideremark#1{\ifvmode\leavevmode\fi\vadjust{\vbox to0pt{\vss
 \hbox to 0pt{\hskip\hsize\hskip1em
 \vbox{\hsize2cm\tiny\raggedright\pretolerance10000
 \noindent #1\hfill}\hss}\vbox to8pt{\vfil}\vss}}}%
\usepackage{color}

\title{\boldmath  Geometric creation of quantum vorticity }

\author[a,b]{Michael R.R. Good}
\author[b,e]{Chi Xiong}
\author[b,c]{Alvin J.K. Chua}
\author[b,d]{Kerson Huang}

\affiliation[a]{Department of Physics, Nazarbayev University, Astana, Kazakhstan}
\affiliation[b]{Institute of Advanced Studies, Nanyang Technological University, Singapore}
\affiliation[c]{Institute of Astronomy, University of Cambridge, Cambridge, UK}
\affiliation[d]{Department of Physics, Massachusetts Institute of Technology, Cambridge, MA, USA}
\affiliation[e]{School of Physical and Mathematical Sciences, Nanyang Technological University, Singapore}

\emailAdd{michael.good@nu.edu.kz}
\emailAdd{xiongchi@ntu.edu.sg}
\emailAdd{ajkc3@ast.cam.ac.uk}
\emailAdd{kerson@mit.edu}

\abstract{We consider superfluidity and quantum vorticity in rotating spacetimes. The system is described by a complex scalar satisfying a nonlinear Klein-Gordon equation. Rotation terms are identified and found to lead to the transfer of angular momentum of the spacetime to the scalar field.  The scalar field responds by rotating, physically behaving as a superfluid, through the creation of quantized vortices. We demonstrate vortex nucleation through numerical simulation. 

}

\begin{document}

\maketitle
\flushbottom


\section{Introduction}

Can spacetime curvature generate vorticity?  Does the superfluid transfer of angular momentum happen on large scales?  It is already known that in at least one example, in fair contrast to liquid helium experiments in the lab, there is the interesting case of the large-scale, high temperature environment of a neutron star, where superfluidity is the leading explanation for pulsar glitches \cite{Alford:2007xm, Lattimer:2004pg}. In this paper, we investigate the question of whether non-trivial spacetime curvature itself may help catalyze superfluid vorticity in a proof-of-principle toy model.  To this end, it is known that the vacuum is not empty but filled with at least one complex scalar field, namely the Higgs field postulated to generate mass in the Standard Model of particle physics. The scalar field's existence has found experimental support in the discovery of the associated field quanta, the Higgs boson \cite{ATLAS, CMS}.  The nonlinear quantum phase dynamics of such a field, on a macroscopic scale, can lead to superfluidity.  There are proposals that associate many cosmological phenomena with this superfluidity, chief among which are dark energy, dark matter, and inflation \cite{Zurek85}-\cite{Huang}.  For instance, it has been shown that Bose-Einstein condensates (BEC), which sustain superfluidity under certain conditions, can be formed in curved spacetime and considered as a candidate for dark matter \cite{Harko}-\cite{Suarez}. Therefore superfluidity may play an important role in the study of dark matter and this gives another motivation to consider superfluidity at cosmological scales. 

In this wide context, we study the superfluidity of the Higgs vacuum in rotating spacetimes. Quantized vorticity is the only means through which a superfluid can rotate, and as such, stands as a prominent signature of superfluidity. 
Steady regularly-distributed vortex lattices have been experimentally observed in rotating superfluid helium \cite{Donelly} and the rotating BEC of cold trapped atoms \cite{Fetter}. This is a very interesting and distinctive macroscopic quantum phenomenon, and one may wonder whether it can happen on astrophysical or cosmological scales. Our observation is that rotating black holes, such as the Kerr black hole in (3+1)-dimensions \cite{Kerr} and the BTZ black hole in (2+1)-dimensions \cite{Banados}, can provide a similar rotating environment as in condensed physics experiments and hence may induce quantum vortices in the cosmic superfluid or the Higgs vacuum. However, the rotating ``bucket'' here, significantly different from the container for liquid helium and the laser trap for BECs, is the spacetime itself. It has never been demonstrated in principle that curved spacetime itself can nucleate a quantum vortex lattice. We emphasize that we do not solve the Kerr and BTZ cases explicitly, but instead, more modestly show that their equations of motion provide the right key ingredients to form superfluid vortices.  


In general relativity, frame-dragging (or the Lense-Thirring effect \cite{Lense}) refers to a special distortion of spacetime geometry caused by a rotating mass. It has been detected \cite{Everitt}, and occurs because the spacetime has angular momentum.  We explore the possibility that the frame-dragging effect of a spacetime with angular momentum can create quantum vortices in the cosmic superfluid.  This is quantum vortex formation induced by angular momentum transfer from the spacetime to the field.  We refer to it as {\it geometrical creation of quantized vorticity}.  

In both the superfluid helium and the BEC cases, the system is described by a macroscopic wave function $\Psi (\vec{x}, t)  = |\Psi(\vec{x}, t) | e^{i \alpha (\vec{x}, t)}$ as an order parameter. $\Psi(\vec{x}, t)$ satisfies the nonlinear Schr\"odinger equation (NLSE) or the Gross-Pitaevskii equation (GPE), which are not suitable for relativistic systems. Therefore we use the nonlinear Klein-Gordon equation (NLKG), promoting the order parameter $\Psi(\vec{x}, t)$ to a complex scalar field $\Phi(\vec{x}, t)$.  We include a usual nonlinear potential (for simplicity we use a Higgs potential in the present paper), but the main difference here is the use of the non-trivial topology of the curved space NLKG equation, which will be subject to the effect of black hole spacetimes. For a comparison of the KG equation with the GPE equation in the context of curved spacetime, see \cite{Castellanos:2013ena}. 

In general, this Klein-Gordon equation should be solved in combination with the Einstein equation; however, we neglect the backreaction of the complex scalar field $\Phi(\vec{x}, t)$ on the stress-energy tensor, and treat the rotating black hole as the spacetime background in the canonical semi-classical approach. Despite neglecting backreaction, it is difficult to find numerical solutions for the NLKG in such backgrounds. We are therefore compelled to make a slow-rotation approximation under which two rotation terms, a Coriolis term and a centrifugal term, emerge at first and second order in the rotational angular velocity $\Omega$, respectively. Based on numerical computations, we find that the first order Coriolis term is the most important. We show that it leads to vortex formation, and therefore that quantum vortices can be generated by the frame-dragging effect of black holes. The fast-rotating case is left for future investigation.  

An interesting finding is that the emergence of the rotation terms occurs for both the BTZ case and the Kerr case.  We show this explicitly by expanding their equations of motion in terms of angular velocity.  We also find a direct link between the BTZ background and a simple rotation metric.  We numerically evolve the field in the (2+1)-dimensional case with appropriate approximations, looking for quantum vorticity.  Vortex nucleation in this case emerges in a similar way to superfluid helium and BEC vortex lattices. We have not solved the more formidable Kerr case.  Even though we have identified the same indispensable rotation terms in the Kerr equations of motion, the Kerr background geometry presents a significantly more challenging numerical endeavor.  Nevertheless, in the Kerr case (like the BTZ case), it is still certain that the Coriolis rotation term will be responsible for the transfer of angular momentum to the field, catalyzing quantum vortex formation.

\section{NLKG in an arbitrary metric}

The occurrence of superfluidity signals the spontaneous breaking of a global symmetry, which is taken to be a $U(1)$ group in the present article. This can be readily described by a complex scalar field theory  $\Phi$ with a Higgs-type potential. 
Therefore the field $\Phi$, not particularly tied to any vacuum field in particle theory, satisfies a nonlinear Klein-Gordon equation (NLKG) in an arbitrary metric (setting $\hbar=1$)
\begin{equation} 
\square\Phi+\lambda(|\Phi|^{2}-F_{0}^{2})\Phi=0,\label{NLKG}
\end{equation}
where 
\begin{equation}
\square\Phi\equiv\left(  -g\right)  ^{-1/2}\partial^{\mu}\left[
\left(  -g\right)  ^{1/2}g_{\mu\nu}\partial^{\nu}\Phi\right], ~~~g \equiv \det (g_{\mu\nu}).
\end{equation}
The metric $g_{\mu\nu}$ is the means through which frame-dragging effects will enter, while
$\lambda$ and $F_{0}$ are the self-interaction constant of the $\Phi$ field and the vacuum expectation value of its modulus, respectively. 
The nonlinear self-interaction spontaneously breaks the global $U(1)$ symmetry, and maintains a nonzero vacuum field
$F_{0}$. As shown in \cite{HXZ}, this vacuum field is stable against collapse due to
self-gravitation, i.e., the equivalent Jeans length is greater than the radius
of the universe.  (See \cite{Bastrukov:2009zz} for a model of a self-gravitating liquid mass which vibrates.)  The NLKG equation is used (as opposed to the non-linear Sch\"odinger equation), because relativistic effects are essential in this context. In addition, the NLKG equation is more generalizable to quantum field theory in curved spacetime, and the relationship between angular velocity and the number of vortices (Feynman's relation) becomes modified in the relativistic regime \cite{XGGLH}.

Note that an NLKG-type equation may also emerge in the formalism of relativistic BEC in flat and curved spacetime.  As shown in refs.  \cite{Fagnocchi:2010sn, Bettoni:2013zma}, one can start with an interacting relativistic scalar boson $\phi$ at finite temperature $T$, and then calculate the difference between the numbers of bosons and anti-bosons and its relation to the critical temperature $T_c$, under which a phase transition may happen, especially when $T \ll T_c$ almost all particles are condensed at the ground state, indicating the formation of a BEC. Separating the collective wave function ($\Phi)$ from the quantum fluctuation ($ \varphi$) as $\phi = \Phi ( 1+ \varphi)$, one finds that the scalar field $\Phi$ satisfies an NLKG-type equation (see refs.  \cite{Fagnocchi:2010sn, Bettoni:2013zma} for details).


In a phase representation $\Phi=Fe^{i\sigma},$ the superfluid velocity is
given by $\mathbf{v}_{\text{s}}=\xi_{\text{s}}\nabla\sigma,\,$\ where
$\xi_{\text{s}}=c^{2}\left(  -\partial\sigma/\partial t\right)  ^{-1}$ is a
spacetime dependent factor that makes $\left\vert \mathbf{v}_{\text{s}
}\right\vert <c$. In the non-relativistic limit we have $\xi_{\text{s}}\rightarrow$
$\hbar/m$, where $m$ is a mass scale \cite{HXZ, XGGLH}. In practice, it is simpler to work
with $\nabla\sigma$. A quantized vortex is a solution to the NLKG with
\begin{equation}
{\displaystyle\oint\limits_{C}}
\nabla\sigma\cdot d\mathbf{s=}2\mathbf{\pi}n\text{ \ }\left(  n=0,\pm
1,\pm2,\ldots\right)  ,
\end{equation}
over some spatial closed loop $C.$ Such quantized vortices have been extensively studied in liquid helium and Bose-Einstein condensates of cold trapped atoms, both experimentally and theoretically \cite{Donelly, Fetter, HXZ, XGGLH}.  We will search for these quantized vortex solutions in the following section.

\section{NLKG in the BTZ metric}

We want to extract specific terms in the NLKG, Eq.~(\ref{NLKG}), that arise from the metric and induce vortex creation. For orientation, we consider first the BTZ metric \cite{Banados} in (2+1)-dimensional spacetime, which is a vacuum solution to Einstein's equation in the region outside of a rotating ``star'' (a central distribution that has both mass and angular momentum). The line element in spatial polar coordinates is given by
\begin{equation}
ds_{\text{B}}^{2}=g_{tt}dt^{2}+g_{rr}dr^{2}+g_{\phi\phi}d\phi^{2}+2g_{t\phi
}dtd\phi\text{,}\label{btzmetric}
\end{equation}
or
\begin{equation} ds_{\text{B}}^{2} = \left(g_{tt} - \frac{g_{t\phi}^2}{g_{\phi\phi}}\right)dt^2 + g_{rr}dr^2 + g_{\phi\phi} \left(d\phi +  \frac{g_{t\phi}}{g_{\phi\phi}} dt \right)^2, \end{equation}
with components in matrix form using $(t,r,\phi)$ coordinates,
\begin{equation}
\label{BTZmetricform}
(g_{\mu\nu})^{\mathrm{BTZ}} = \left( 
\begin{array}{ccc}
8 G M - \frac{c^2 r^2}{a^2} & 0 & -\frac{4 G J}{c^2} \\ 
0 & \frac{1}{\frac{r^2}{a^2}+\frac{16 G^2 J^2}{c^6 r^2}-\frac{8 G M}{c^2}} & 0 \\ 
-\frac{4 G J}{c^2} & 0 & r^2
\end{array} \right),
\end{equation}
where $M$ is the black hole mass, $J$ is the angular momentum of the spacetime and $a$ is the AdS radius. The AdS radius characterizes the maximally symmetric Lorentzian manifold with constant negative scalar curvature, the analogue of hyperbolic space, in the same way that the radius of de Sitter space is the radius of a sphere since de Sitter space is the analogue of spherical space. This AdS radius is a radius of curvature of a generalized sphere in the sense that it is a collection of points for which the `distance', as determined by quadratic form, from the origin is constant.  The BTZ black hole is a solution for (2+1)-dimensional gravity with a negative cosmological constant, i.e. asymptotically AdS. It is used as a toy model to illustrate relevant physical content.

The frame-dragging component $g_{t\phi}$ implies a local rotational angular velocity
\begin{equation}  \label{Omega_B}
\Omega_{\text{B}}=-\frac{g_{t\phi}}{g_{\phi\phi}}=\frac{4GJ}{c^{2}r^{2}},
\end{equation}
which is independent of the mass $M.$ The NLKG equation in the BTZ metric is Eq.~(\ref{NLKG}) with $g_{\mu\nu}$ given by Eq.~(\ref{BTZmetricform}),\footnote{Here we temporarily take the units that are usual for the BTZ metric, $c=1$ and $G = 1/8$, to illustrate the simplicity of the hypergeometric result.} i.e., 
\begin{eqnarray}  \label{btzeom}
\Box \Phi &=& - g_{rr}  \bigg(\frac{\partial^2 \Phi}{\partial t^2} + \frac{J}{r^2} \frac{\partial^2 \Phi}{\partial t \partial \phi}  + \frac{J^2}{4 r^4} ~\frac{\partial^2 \Phi}{\partial \phi^2} \bigg)  + \frac{1}{r^2}\frac{\partial^2 \Phi}{\partial \phi^2} + \frac{1}{r}\frac{\partial}{\partial r} \bigg( \frac{r}{g_{rr}} \frac{\partial \Phi}{\partial r} \bigg)  \cr
&=&  - \lambda(|\Phi|^{2}-F_{0}^{2})\Phi ,
\end{eqnarray}
where $g_{rr}$ is given by
\begin{equation}
g_{rr} = \frac{1}{\frac{J^2}{4 r^2} + \frac{r^2}{a^2} - M} = \frac{4 a^2 r^2}{4 r^4 + a^2 (J^2 - 4 M r^2)}.
\end{equation}

To illustrate the structure of Eq.~(\ref{btzeom}), we apply the usual method of separation of variables:
\begin{equation} \label{ansatz}
\Phi(t, r, \phi) = ~e^{-i \omega t} ~ e^{im \phi} ~R(r)
\end{equation}
to obtain
\begin{equation}  \label{eomR}
\bigg[ g_{rr} \bigg( \omega - m \Omega_B  \bigg)^2 - \frac{m^2}{r^2} + \frac{1}{r}\frac{d}{dr} \bigg( \frac{r}{g_{rr}} \frac{d}{d r} \bigg) \bigg] R(r) = \lambda ( R^2 - F_0^2) R(r).
\end{equation}
Interestingly, the corresponding homogeneous differential equation of Eq.~(\ref{eomR}) (i.e. removing the nonlinear potential term on the right-hand side of Eq.~(\ref{eomR})) is actually a hypergeometric equation \cite{KKS}. However, the ansatz Eq.~(\ref{ansatz}) cannot lead to a vortex lattice solution and we have to solve the equation Eq.~(\ref{btzeom}) without it.

Reinstating $G$ and $c$, we now show that under an appropriate approximation, solving Eq.~(\ref{btzeom}) is equivalent to solving the NLKG in the following comparison metric that describes rotation in flat (2+1)-dimensional spacetime, with angular velocity $\Omega_{0}$:  
\begin{equation}
ds_{\text{comparison}}^{2}=-c^{2}dt^{2}+dr^{2}+r^{2}(d\phi-\Omega_{0}dt)^{2}.
\label{comp}
\end{equation}
The effect of rotation enters the NLKG only through $\square\Phi$, which separates into a non-rotational term plus terms that can be identified with the Coriolis force and centrifugal force:
\begin{equation} \label{compeom}
\square\Phi=\square^{(0)}\Phi+R_{\text{Coriolis}}+R_{\text{centrifugal}},
\end{equation}
where $\square^{(0)}\Phi$ is that for the (2+1)-dimensional Minkowski metric, 
\begin{equation} \label{NRB} \square^{(0)} \Phi ( t, r, \phi ) = -\frac{1}{c^2}\frac{\partial ^2\Phi (t,r,\phi )}{\partial t^2}+\frac{\partial ^2\Phi (t,r,\phi )}{\partial r^2}+\frac{1}{r}\frac{\partial \Phi (t,r,\phi )}{\partial r} +\frac{1}{r^2}\frac{\partial ^2\Phi (t,r,\phi )}{\partial \phi ^2},  \end{equation}
and the rotational terms are 
\begin{equation}
R_{\text{Coriolis}}=-\frac{2\Omega_{0}}{c^{2}}\frac{\partial^{2}\Phi}{\partial
t\partial\phi},\text{ \ \ }R_{\text{centrifugal}}=-\frac{\Omega_{0}^{2}}%
{c^{2}}\frac{\partial^{2}\Phi}{\partial\phi^{2}}.\text{\ \ } \label{rot}%
\end{equation}

The equivalence between the two approaches comes from the fact that the BTZ metric is formally connected to the comparison metric for the special values 

\begin{equation}
M=-\frac{c^{2}}{8G},\quad a^{2}=-\frac{c^{2}}{\Omega_{0}^{2}}, \end{equation}  
which leads to 
\begin{equation}
g_{rr} = \frac{1}{1+ \frac{r^2}{c^2} ( \Omega_B^2 - \Omega_0^2)} = 1 - \frac{r^2}{c^2} ( \Omega_B^2 - \Omega_0^2) + \mathcal{O} ((\Omega_B^2 - \Omega_0^2)^2),
\end{equation}
and hence 
\begin{equation} 
(g_{\mu\nu})^{\mathrm{BTZ}} - (g_{\mu\nu})^{\textrm{\tiny{Comparison}}}  = \left( 
\begin{array}{ccc}
0 & 0 &  (\Omega_0 - \Omega_B) r^2  \\ 
0 & - \frac{r^2}{c^2} ( \Omega_B^2 - \Omega_0^2) & 0 \\ 
 (\Omega_0 - \Omega_B) r^2 & 0 & 0
\end{array} \right) +  \mathcal{O} ((\Omega_B^2 - \Omega_0^2)^2).\label{equiv}
\end{equation}
Therefore the difference between these two metrics vanishes if we assume that $\Omega_B$ is approximately constant: $\Omega_B \approx \Omega_0$.  Note that from Eqn. (\ref{Omega_B}) $\Omega_B = 4 G J/(c^2 r^2) $ can be expanded around a relatively large distance $r_0$
\begin{equation}
\Omega_B = \frac{4 G J}{c^2 r^2} = \frac{4 G J}{c^2} \left( \frac{1}{r_0^2} - \frac{2}{r_0^3} \Delta r + \mathcal{O} ( (\Delta r)^2 \right),
\end{equation}
therefore in the ring region $r_0 - \Delta r < r < r_0 + \Delta r $, one can consider the angular velocity $\Omega_b \approx 4 G J/ (c^2 r^2_0)$ and call this quantity $\Omega_0$. Whether this approximation is legitimate depends on the scale of the physical phenomenon to be studied. In this paper we are interested in quantum vortices whose scale is determined by their core size, which is certainly very small compared to the macroscopic scale of the blackhole geometry. 

Under this approximation, it is easy to read Eq.~(\ref{compeom}), Eq.~(\ref{NRB}) and Eq.~(\ref{rot})  from Eq.~(\ref{btzeom}). (Alternatively, one can simply write the NLKG in the comparison metric to identify the terms $R_{\text{Coriolis}}$ and $ R_{\text{centrifugal}}$ in Eq.~(\ref{rot}) \cite{XGGLH}.) 

In general, we can identify the effective Coriolis force and centrifugal force arising from the
BTZ metric by expanding Eq.~(\ref{btzeom}) in powers of $J$ (or equivalently
$\Omega_{\text{B}}$) to second order, which gives:%
\begin{equation}
R_{\text{Coriolis}}^{\text{B}}=\frac{2\Omega_{\text{B}}}{g_{tt}}\frac
{\partial^{2}\Phi}{\partial t\partial\phi},\text{ \ \ }R_{\text{centrifugal}%
}^{\text{B}}=\frac{\Omega_{\text{B}}^{2}}{g_{tt}}\frac{\partial^{2}\Phi
}{\partial\phi^{2}}.\text{\ } \label{btzrot}%
\end{equation}
\newline There are of course higher order terms in the BTZ wave operator $\square\Phi$; however, the
presence of the terms Eq. (\ref{btzrot}) indicates that the field $\Phi$ does, in fact, feel the rotation
of the star through spacetime geometry. This is to be expected, since
frame-dragging is spacetime dependent. The terms in Eq.~(\ref{btzrot}) depend on $r$ because
$\Omega_{\text{B}}$ and $g_{tt}$ depend on $r$, and they approach the form
Eq.~(\ref{rot}) when $g_{tt}\rightarrow-c^{2}$, in the limits:
\begin{equation} 
\label{BTZapprox} (g_{tt})^{\mathrm{BTZ}}\left|^{M=-\frac{c^2}{8G}}_{a \rightarrow \infty} \right. = -c^2,
						\quad (g_{tt})^{\mathrm{BTZ}}\left|^{M\rightarrow 0}_{r \rightarrow a} \right. = -c^2.
\end{equation}
It is noted that in the first limit of Eq.~\eqref{BTZapprox}, (after $M$ is set to its negative value), $a\rightarrow \infty$ is equivalent to the regime $r \Omega \ll c$ in the context of the formal reduction $a^2 = -c^2/\Omega^2$.  In addition, when this first limit of Eq.~\eqref{BTZapprox} and $J \rightarrow 0$ is applied to the BTZ metric Eq.~\eqref{BTZmetricform}, the result is the flat and non-rotating (2+1)-dimensional Minkowski metric.  The second limit of Eq.~\eqref{BTZapprox}, where $M\rightarrow0$, allows one to obtain the empty space vacuum state. The physical significance of the first limit lies in its consistency of approach to flat spacetime.  The physical significance of the second limit lies in the fact that it furnishes a vanishing, non-negative black hole mass, approaching the vacuum state of empty space (not AdS space).  In the context of AdS space emerging as a bound state from a continuous black hole mass spectrum (after one has set $M=-c^2/8G$ and $J=0$ but before one has taken the limit $a \rightarrow \infty$) \cite{Banados}, the first limit can be considered a more natural choice.

Although the rotational effects are contained in $\square\Phi$, the nonlinear self-interaction will be necessary to produce vortices. Without the nonlinear potential in Eq. (\ref{NLKG}), the situation is that of a free test particle in the spacetime.  To see this, consider the radial field equation Eq. (\ref{eomR}). The radial field describes some material in the spacetime, while the nonlinear potential brings in self-interaction. The situation here is very similar to that for the NLKG with a rotating star as the source (discussed in \cite{HXZ, XGGLH}); however, the numerical computation in the BTZ case is more challenging, owing to its curved-space nature and the cross derivative in the Coriolis force. 

\begin{figure}
[ptb]
\begin{center}
\includegraphics[
height=1.7391in,
width=3.4765in
]%
{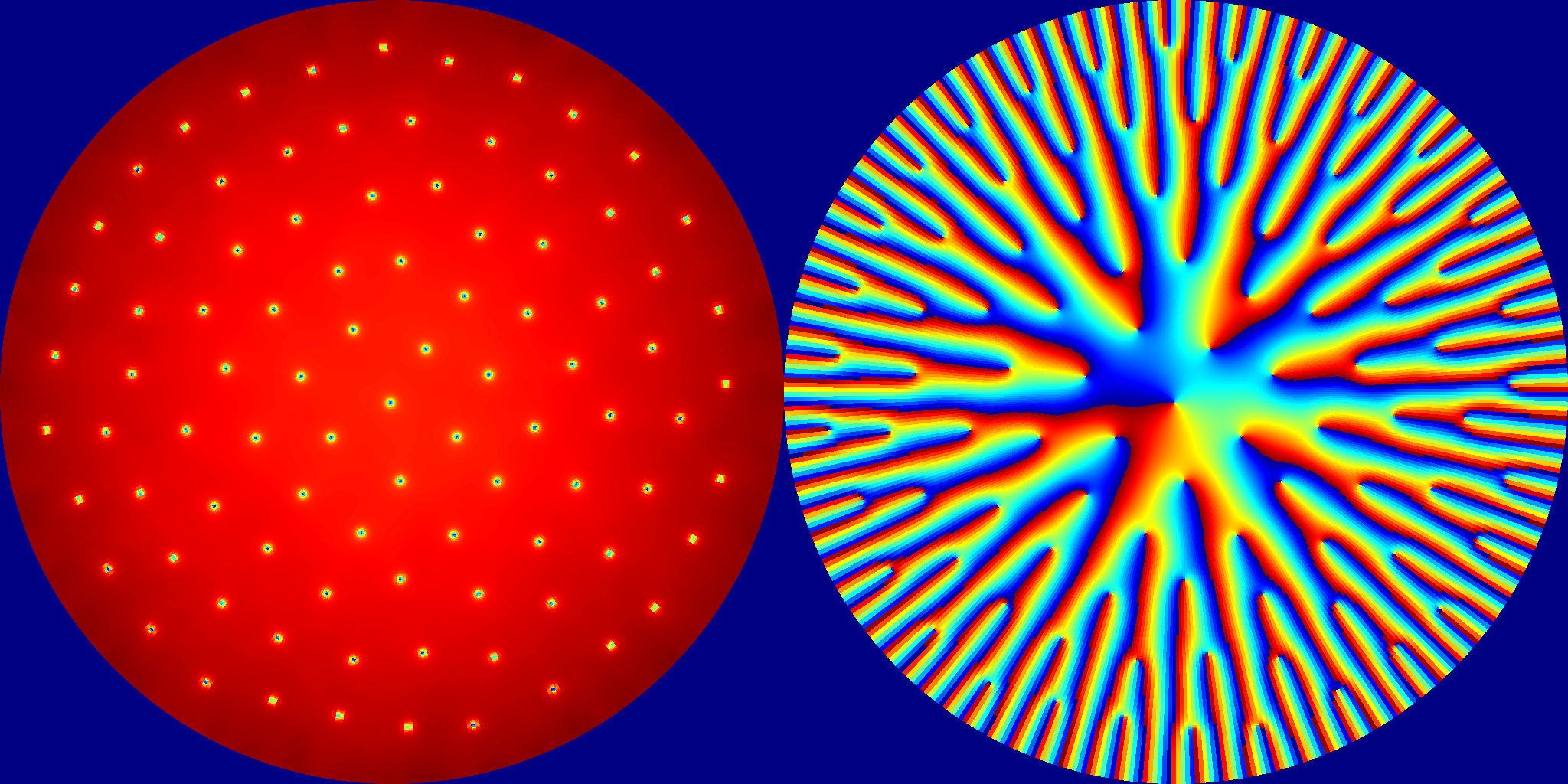}%
\caption{Left panel: Lattice of 100 quantized vortices. This is a contour plot
of the modulus of a complex scalar field that satisfies the nonlinear
Klein-Gordon equation in (2+1)-dimensional curved space-time, with
frame-dragging. Right panel: Contour plot of the phase of the complex scalar
field, showing the ``strings'' across which the phase jumps by 2$\pi$.
The computation is performed in a circular domain with radius $R =120$, and the normalized parameters are $\Omega_0 = 1/240$, $\lambda = 6.4$ and $F_0 =1$. 
}%
\label{fig1}
\end{center}
\end{figure}

To explicitly demonstrate geometric vortex creation, we solve the NLKG numerically in the comparison spacetime, (i.e. Eq. (\ref{NLKG}) with the metric Eq. (\ref{comp})), which retains the frame-dragging rotational terms Eq. (\ref{rot}).  The equivalence to the BTZ metric can be seen from Eq. (\ref{equiv}). This approach is possible to handle numerically and the results can be seen in Fig. (\ref{fig1}). For various choices of model parameters (which we specify), a vortex lattice emerges in the superfluid. 

As we have mentioned earlier in this section (see discussions below Eqn. (\ref{equiv})), for a finite region in space whose size is very small compared with the AdS radius of the BTZ metric, the angular velocity $\Omega_B$ can be considered as approximately constant, $\Omega_0$. The numerical results for the NLKG in the comparison metric are equivalent to those for the NLKG in the BTZ metric under such an approximation.  In Fig. (\ref{fig1}) we used a finite-difference scheme and spectrum methods to decouple the cross derivatives $\partial_t$ and $\partial_\phi$ in the Coriolis term, and a semi-implicit scheme similar to the Crank-Nicolson method for the linear terms and an explicit scheme for the nonlinear term. The boundary of the computation domain is chosen to satisfy $\Omega r \ll c$, and a second order Neumann boundary condition is imposed. The initial condition is an energetic, unstable vortex with high winding number, plus minor perturbation. To reach the ground state, we filter out the high-frequency fluctuations until the field configuration mainly consists of low-frequency components, signaled by the appearance of vortices. The final state is stable and is considered the ground state. For an interesting treatment on how non-minimally coupled free scalar fields are unstable
in the spacetime of compact objects see \cite{Mendes:2014vna}.

Fig. (\ref{fig1}) contains the vortex lattice with a plot of the modulus of the field $|\Phi|$ on the left side and a plot of the phase $\sigma$ on the right side. The locations of vortices are indicated by the dots (i.e. field zeros) in the modulus plot, and by the phase discontinuities around which the color changes from blue to red (see online color version), corresponding to a phase change of $2\pi$. These results demonstrate quantum vortices.  Their distribution is determined by the Coriolis term. As the computation is performed under the slow-rotation approximation $\Omega \ll 1$, the centrifugal term of order $\Omega^2$ causes only a small non-uniformity to the vortex lattice.
Again notice that in the BTZ metric the distribution of vortices in the whole space might not be uniform since the equivalence of the BTZ metric and the comparison metric is only established under the approximation $\Omega_B \approx \Omega_0$ in a small ring region in the BTZ background. However, this region is large enough to accommodate probably billions of quantum vortices.

\section{NLKG in the Kerr metric}

It is now interesting to turn to the Kerr metric\footnote{The Kerr metric may also be parametrized by the Schwarzschild surface gravity $g$ and spring constant $k\equiv M \Omega_K^2$, in order to easily demonstrate its Hawking temperature $2\pi T = g - k$; see \cite{Good:2014uja}.} in (3+1)-dimensional space-time, which describes the region outside of a rotating star
of mass $M$ and angular momentum $J$ \cite{Kerr}. The Kerr metric reduces to the Schwarzschild metric
when $J=0$. The line element is, in spatial spherical coordinates,%
\begin{equation}
ds_{\text{K}}^{2}=g_{tt}dt^{2}+g_{rr}dr^{2}+g_{\theta\theta}d\theta
^{2}+g_{\phi\phi}d\phi^{2}+2g_{t\phi}dtd\phi\text{,}%
\end{equation}
with%
\begin{align}
g_{tt}  &  =-c^{2}\left(  \Delta-\alpha^{2}\sin^{2}\theta\right)  \Sigma
^{-1},\text{ \ \ }g_{rr}=\Sigma\Delta^{-1},\text{ }\nonumber\\
\text{\ }g_{\theta\theta}  &  =\Sigma,\text{ \ }g_{\phi\phi}=-\sin^{2}%
\theta\left[  \alpha^{2}\Delta\sin^{2}\theta-\left(  r^{2}+\alpha^{2}\right)
^{2}\right]  \Sigma^{-1},\nonumber\\
\text{\ }g_{t\phi}  &  =-c\alpha\sin^{2}\theta\left(  r^{2}+\alpha^{2}%
-\Delta\right)  \Sigma^{-1},
\end{align}
where%
\begin{align}
\Delta &  \equiv\alpha^{2}+r^{2}-rr_{s},\text{ \ }\Sigma\equiv\alpha^{2}%
\cos^{2}\theta+r^{2},\text{ \ }\nonumber\\
r_{s}  &  \equiv2GMc^{-2},\text{ \ }\alpha\equiv J/Mc.
\end{align}
The local angular velocity is given by \cite{Wald}%
\begin{equation}
\Omega_{\text{K}}=-\frac{g_{t\phi}}{g_{\phi\phi}}=\frac{\alpha(r^{2}%
+\alpha^{2}-\Delta)c}{(r^{2}+\alpha^{2})^{2}-\Delta\alpha^{2}\sin^{2}\theta}.
\end{equation}

For a comparison metric in (3+1) dimensions, the Coriolis and centrifugal terms have the
same form as in Eq. (\ref{rot}), except that $\phi$ is to be
identified with the azimuthal angle in three spatial dimensions. We calculate $\square\Phi$
using the Kerr metric and write the NLKG Eq. (\ref{NLKG}) explicitly as
\begin{eqnarray}  \label{NLKG_Kerr}
\Sigma \, \square \Phi &=& \left[ -\frac{(r^2+\alpha^2)^2}{\Delta} +\alpha^2 \sin^2\theta \right] ~ \frac{\partial^2 \Phi}{\partial t^2} 
 - \frac{4 \alpha M r}{\Delta}  ~\frac{\partial^2 \Phi}{\partial t \partial \phi} 
 + \frac{\Sigma -2 M r }{\Delta \sin^2 \theta} ~\frac{\partial^2 \Phi}{\partial \phi^2}  \cr
&& + \frac{\partial}{\partial r} \left(\Delta \frac{\partial \Phi}{\partial r} \right) 
 + \frac{1}{\sin \theta}\frac{\partial}{\partial \theta} \left(\sin \theta \frac{\partial \Phi}{\partial \theta} \right)  \cr
&=& -  \Sigma \, [ \lambda(|\Phi|^{2}-F_{0}^{2})\Phi ],
\end{eqnarray}
where an overall factor of $\Sigma$ is included to simplify the expression. Similar to the BTZ case, we extract the Coriolis and centrifugal terms by expanding in powers of $\Omega_{\text{K}}$, to second order:
\begin{equation}
\label{Kerrrot}
R_{\text{Coriolis}}^{\text{K}}=\frac{2\Omega_{\text{K}}}{g_{tt}}\frac
{\partial^{2}\Phi}{\partial t\partial\phi},\text{ \ }R_{\text{centrifugal}%
}^{\text{K}}=\frac{\Omega_{\text{K}}^{2}}{g_{tt}}\frac{\partial^{2}\Phi
}{\partial\phi^{2}}.
\end{equation}
These are formally the same as in the BTZ metric. We recover Eq.~(\ref{rot})
when $g_{tt}\rightarrow-c^{2}$, in the limit $r\gg r_{s}$ and $c\rightarrow
\infty$. These quantities vanish for the Schwarzschild metric, for which
$\Omega_{\text{K}}=0.$ In our calculations, we use the potential approximation
to the unperturbed Schwarzschild metric \cite{MTW}. 

In the slow-rotation regime, once we extract the terms $R_{\text{Coriolis}}$ and $R_{\text{centrifugal}}$ from the Kerr black hole geometry to describe the rotation, the confining effect of the Kerr black hole on the superfluid is not very different from the case of the non-rotating Schwarzschild black hole. It is easy to see that Eq.~(\ref{NLKG_Kerr}) reduces to a NLKG in the Schwarzschild black hole background by setting $a=0$; then, under the Newtonian approximation, an effective confining potential $ \sim - \frac{M}{r}$ appears. Therefore, if $M$ is large enough, rotating black holes can hold some amount of cosmic superfluid through their gravitational pull, and the effective potential plays the same role as the trapping potential in the rotating BEC context. We have not been able to solve Eq.~(\ref{NLKG_Kerr}) exactly. Nevertheless, a numerical computation based on the NLKG with a current-current interaction has been performed in Ref. \cite{XGGLH}.  This interaction (see Eqs. (45 -- 51) in Ref. \cite{XGGLH}), effectively contains a position-dependent angular velocity term and is used to simulate the local rotation terms $R_{\text{Coriolis}}$ and $R_{\text{centrifugal}}$ of the Kerr black hole.  The simulation only focuses on the position-dependent feature of the rotating terms, neglecting the other curved spacetime factors.  However, a solution of a three-dimensional vortex-ring lattice was found (see Fig. 4 in Ref. \cite{XGGLH}), and it will be interesting to see whether such a solution exists for Eq. (\ref{NLKG_Kerr}).

\section{Conclusions and Outlooks}

In this paper we have studied, from a mathematical point of view, the nonlinear Klein-Gordon equation in the (2+1)-dimensional BTZ black hole background and the (3+1)-dimensional Kerr black hole background. Physically speaking, we have studied the behavior of superfluids driven by rotating black holes (BTZ and Kerr) via the frame-dragging effect, and observed the formation of quantum vortex lattices by numerically solving the NLKG equation under tractable approximations (slow and constant rotation). In the case of slow rotation, we have identified Eq.~(\ref{rot}), Eq.~(\ref{btzrot}), and Eq.~(\ref{Kerrrot}) as the relevant rotation terms for studying quantum vorticity in rotating backgrounds. The common form taken by these terms is explicitly shown and discussed. In addition, the vortex lattice in Fig. (\ref{fig1}) has the same pattern as the lattices seen in rotating superfluid helium and rotating BECs \cite{Donelly, Fetter}. Although our results are derived in a restricted region of parameter space, they show how angular momentum in the metric can be transferred to the fields, resulting in vorticity and hence confirming superfluidity.

We do not speculate on whether quantum vorticity should be manifest in astrophysical observations. Our results confirm, only in principle, that rotating geometry induces vorticity. Our purpose here was modest: to see if vorticity could be sourced by the angular momentum in the spacetime. As this has been confirmed, we will however remark that nucleation of quantum vortices in an astrophysical context could require significant frame-dragging, and this may be the case in the neighborhood of a black hole with ultra-high angular momentum. It has been speculated elsewhere that the so-called ``non-thermal filaments'' observed near the center of the Milky Way are due to quantized vorticity \cite{HLT2}.

As a related issue, it is interesting to mention that as the angular momentum increases, one expects that vortex filaments would
appear closer to the black-hole surface, forming a hydrodynamic
boundary layer; this may be relevant to the problem of black-hole
collapse. It is well known in hydrodynamics that such a boundary layer can
accommodate any necessary boundary conditions for the surrounding fluid. One
of the outstanding theoretical problems in general relativity is to extend the
Oppenheimer-Snyder solution \cite{OS} of black-hole collapse in the Schwarzschild
metric, which carries no angular momentum, to the Kerr metric. A main obstacle
appears to be the lack of an appropriate generalization of the interior
metric, which in the Oppenheimer-Snyder case is just the non-rotating closed
Robertson-Walker metric. It is therefore tempting to contemplate that when a hydrodynamic boundary layer is present,
the outside Kerr metric could be joined onto any interior metric.

The relativistic character of the curved-space NLKG equation is appropriate for addressing questions involving strong gravitational fields, such as those in the vicinity of the curved spacetime of a rotating black hole. Since quantum vorticity is synonymous with superfluids and is indicative of dynamical complex fields and non-linear pattern-forming systems, its presence in a gravitational context is particularly welcome. We hope that these investigations contribute to the productive and fascinating enterprise that has been the consideration of gravitational influence on quantum fields.

\section*{Acknowledgements}
We thank Weizhu Bao and Yong Zhang of the National University of
Singapore for helpful discussions and suggestions on numerical computations,
and Yulong Guo for assistance in much of the coding. CX is supported by the research grant from the Institute of Advanced studies, Nanyang Technological University, Singapore and MRRG is funded in part by the Social Policy Grant at Nazarbayev University.


\begin{thebibliography}{99}
\bibitem{Alford:2007xm} 
  M.~G.~Alford, A.~Schmitt, K.~Rajagopal and T.~Sch\"{a}fer,
  Rev.\ Mod.\ Phys.\  {\bf 80}, 1455 (2008)
  [arXiv:0709.4635 [hep-ph]].

\bibitem{Lattimer:2004pg} 
  J.~M.~Lattimer and M.~Prakash,
  Science {\bf 304}, 536 (2004)
  [astro-ph/0405262].

\bibitem{ATLAS}
ATLAS collab., \textit{Phys. Letters} B \textbf{716}, 1 (2012).

\bibitem{CMS}
CMS collab., \textit{Phys. Letters} B \textbf{716}, 30 (2012).

\bibitem{Zurek85} 
  W.~H.~Zurek,
  Nature {\bf 317}, 505 (1985).

\bibitem{Sin92} 
  S.~J.~Sin,
  Phys.\ Rev.\ D {\bf 50}, 3650 (1994)
  [hep-ph/9205208].

\bibitem{Volovik96} 
  G.~E.~Volovik,
  Czech.\ J.\ Phys.\  {\bf 46}, 3048 (1996)
  [cond-mat/9607212].

\bibitem{Volovik01} 
  G.~E.~Volovik,
  Phys.\ Rept.\  {\bf 351}, 195 (2001).

\bibitem{HLT1}
K. Huang, H. B. Low, and R. S. Tung, \textit{Class. Quantum Grav}.
\textbf{29}, 155014 (2012); arXiv:1106.5282.

\bibitem{HLT2}
K. Huang, H. B. Low, and R. S. Tung, \textit{Int. J. Mod. Phys.} \textbf{A}
27, 1250154 (2012); arXiv:1106.5283.

\bibitem{Harko}
C. G. B\"{o}ehmer and T. Harko, \textit{J. Cosmol. Astropart. Phys.}, \textbf{06}, 025 (2007).

\bibitem{JWLee09}
J.-W. Lee, \textit{J. Korean Phys. Soc}. \textbf{54}, 2622 (2009). 

\bibitem{Fagnocchi:2010sn} 
  S.~Fagnocchi, S.~Finazzi, S.~Liberati, M.~Kormos and A.~Trombettoni,
  New J.\ Phys.\  {\bf 12}, 095012 (2010)
  [arXiv:1001.1044 [gr-qc]].

\bibitem{Bettoni:2013zma} 
  D.~Bettoni, M.~Colombo and S.~Liberati,
  JCAP {\bf 1402}, 004 (2014)
  [arXiv:1310.3753 [astro-ph.CO]].

\bibitem{Popov1}
V. Popov, \textit{Phys. Lett.} B \textbf{686}, 211 (2010).

\bibitem{HYLing1}
B. Kain and H.Y. Ling, \textit{Phys. Rev}. D\textbf{82}, 064042 (2010).

\bibitem{Suarez}
A. Suarez, V.H. Robles, and T. Matos, Astrophys.\ Space Sci.\ Proc.\  {\bf 38}, 107 (2014)
  [arXiv:1302.0903 [astro-ph.CO]].

\bibitem{HXZ}
K. Huang, C. Xiong, and X. Zhao, \textit{Int. J. Mod. Phys.} \textbf{A} 29,
1450014 (2014); arXiv:1304.1595.

\bibitem{Huang}
K. Huang, \textit{Int. J. Mod. Phys.} \textbf{A} 28, 1330049 (2013); arXiv:1309.5707.


\bibitem{Donelly}
R. J. Donelly, \textit{Quantized vortices in helium II} (Cambridge University Press, 1991).

\bibitem{Fetter}
For a review, see e.g. A. L. Fetter,\textit{ Rev. Mod. Phys}. \textbf{81}, 647 (2009).

\bibitem{Kerr}
R. P. Kerr, \textit{Phys. Rev.~Lett.~}\textbf{11},237 (1963).

\bibitem{Banados}
M. Ba\~{n}ados, C. T{eitelboim,~}a{nd~}J. {Zanelli,}~{ \textit{Phys.
Rev.~Lett. }}\textbf{69}, 1849 (1992).


\bibitem{Lense}
J. Lense and H. Thirring, \textit{Phys. Zeits}. \textbf{19}, 156 (1918).

\bibitem{Everitt}
C. Everitt \textit{et. al}., \textit{Phys. Rev. Lett.}, 106, 221101
(2011); arXiv:1105.3456.

\bibitem{Castellanos:2013ena} 
  E.~Castellanos, C.~Escamilla-Rivera, A.~Macias and D.~Nunez,
  JCAP {\bf 1411}, no. 11, 034 (2014)
  doi:10.1088/1475-7516/2014/11/034
  [arXiv:1310.3319 [gr-qc]].

\bibitem{Bastrukov:2009zz} 
  S.~I.~Bastrukov, H.~K.~Chang, E.~H.~Wu and I.~V.~Molodtsova,
  Mod.\ Phys.\ Lett.\ A {\bf 24}, 3257 (2009).
  doi:10.1142/S0217732309032137

\bibitem{XGGLH} 
 C.~Xiong, M.~R.~R.~Good, Y.~Guo, X.~Liu and K.~Huang, \textit{Phys. Rev.} \textbf{12}, D90, 125019 (2014).
  arXiv:1408.0779 [hep-th].
	
\bibitem{KKS}
M. Kenmoku, M. Kuwata, K. Shigemoto, Class.Quant.Grav.25:145016, (2008).

\bibitem{Good:2014uja} 
  M.~R.~R.~Good and Y.~C.~Ong,
  Phys.\ Rev.\ D {\bf 91}, no. 4, 044031 (2015)
  [arXiv:1412.5432 [gr-qc]].
	
\bibitem{Mendes:2014vna} 
  R.~F.~P.~Mendes, G.~E.~A.~Matsas and D.~A.~T.~Vanzella,
  Phys.\ Rev.\ D {\bf 90}, no. 4, 044053 (2014)
  doi:10.1103/PhysRevD.90.044053
  [arXiv:1407.6405 [gr-qc]].
	
\bibitem{Wald}
R. M. Wald, \textit{General Relativity} (University of Chicago Press, 1984), p.319.

\bibitem{MTW}
C. W. Misner, K. S. Thorne, and J. A. Wheeler, \textit{Gravitation}
(Freeman, San Francisco, 1973), Chap.25, p.636.

\bibitem{OS}
J. R. Oppenheimer and H. Snyder, \textit{Phys.~Rev}{.}~\textbf{56}, 455 (1939).


 \end{thebibliography}
\end{document}